# Privacy Protection and Video Manipulation in Immersive Media

LESLIE WÖHLER, The University of Tokyo, JSPS Research Fellow, Japan

SATOSHI IKEHATA, National Institute of Informatics, Japan and Tokyo Institute of Technology, Japan

KIYOHARU AIZAWA, The University of Tokyo, Japan

In comparison to traditional footage, 360° videos can convey engaging, immersive experiences and even be utilized to create interactive virtual environments. Like regular recordings, these videos need to consider the privacy of recorded people and could be targets for video manipulations. However, due to their properties like enhanced presence, the effects on users might differ from traditional, non-immersive content. Therefore, we are interested in how changes of real-world footage like adding privacy protection or applying video manipulations could mitigate or introduce harm in the resulting immersive media.

## 1 MOTIVATION

Recently, the creation of 360° videos has become easily feasible for non-professional users, opening countless new possibilities as they can not only evoke a high sense of realism, presence, and engagement [1] but also be used to create realistic virtual environments [4, 5].

Despite their positive properties, **the usage of real-world footage in immersive environments also brings about many new challenges to avoid potential harm**. One area of interest is the *privacy protection of bystanders*. Oftentimes, bystanders do not feel comfortable about being recorded and worry about the usage of their video data [2]. While visible cameras lead to decreased concerns as people can choose to avoid them [3], this is not easily possible for omnidirectional recordings emphasizing the need of reliable anonymization. Furthermore, with the rise of AI technology, highly realistic video editing has become possible opening many new creative directions but also enabling potential *harmful manipulations*. While the perception of manipulations like face-swapping have been investigated for regular videos [9], the effects in onmidirectional videos could differ from non-immersive experiences as the high sense of realism might make viewers more likely to believe in the authenticity of the content.

Therefore, **we believe that it is essential to investigate and understand the specific effects of video editing and manipulations in immersive media to mitigate harms and create ethical content**.

## 2　PREVIOUS EXPERIENCE AND POSITION OF THE AUTHORS

**We are working towards the creation of virtual environments from real-world 360° videos** to enable the seamless and immersive exploration of full city areas either directly based on the video content [4] or by adding 3D avatars to facilitate multi-user interaction [5]. Our frameworks aim to achieve automatic generation from real-world data so that photo-realistic virtual environments can be realized even by novice users without knowledge of 3D modeling.

One of the challenges regarding virtual worlds from real-world data is the need to anonymize facial identities. While techniques like covering or blurring of the facial area are typically used, these visibly change the video content and could potentially reduce the positive aspects of 360° videos like their high realism and presence. Therefore, **we are currently investigating the perception of facial anonymization techniques in 360° videos of public spaces**. Next to traditional facial anonymization techniques, we also analyze the use of face-swapping, a video editing technique that can be used to exchange the facial appearance of people in videos. Based on our research on regular videos, face-swapping is highly realistic, visually appealing, and can retain the original facial expressions [6–9] which suggest that it could be a valuable alternative for facial anonymization. Preliminary results of our experiments indicate that *face-swapping is difficult to recognize even in 360° videos, highlighting the potential harm of the technique when used to create malicious modifications in immersive media*. Despite this, not only the traditional techniques, but also face-swapping, reduced the perceived realism of the 360° videos as well as the presence of viewers when the content is viewed with a head-mounted display. *Therefore, our results indicate a loss of presence for anonymized videos which could lead to conflict between the need for privacy protection and the interests of creators to enhance the presence in immersive media*.

As the automatic nature of our framework can allow anyone to create their own environments from 360° videos, **we aim to further our understanding of possible harms to ensure the ethically responsible usage of our technology**. Additionally, we would like to contribute to the development of mitigating strategies to create pleasant and safe immersive environments.